\documentclass[aps,prd,onecolumn,superscriptaddress,preprintnumbers,showpacs,amsmath,amssymb]{revtex4}
\usepackage{epsf}
\usepackage{latexsym}
\usepackage{amsmath}
\usepackage{amsfonts}
\usepackage{amssymb}
\usepackage{graphicx}
\newcommand{\be}{\begin{eqnarray}}
\newcommand{\ee}{\end{eqnarray}}
\newcommand{\ud}{\mathrm{d}}

\newcommand{\lp}{\ell_{\rm p}}
\newcommand{\mpl}{M_{\rm p}}

\newcommand{\mdd}{M_{(5)}}
\newcommand{\md}{M_{(D)}}

\newcommand{\ld}{\ell_{(D)}}
\newcommand{\mew}{M_{\rm ew}}
\newcommand{\meff}{M_{\rm eff}}

\newcommand{\geff}{g_{\rm eff}}
\newcommand{\mc}{M_{\rm c}}

\newcommand{\rh}{R_{\rm H}}

\begin{document}                                                                
\title{Microcanonical description of (micro) black holes}
\author{R.~Casadio}
\email{casadio@bo.infn.it}
\affiliation{Dipartimento di Fisica, Universit\`a di Bologna, via Irnerio 46, 40126 Bologna, Italy}
\affiliation{INFN, Sezione di Bologna, Via Irnerio 46, I-40126 Bologna, Italy}
\author{B. Harms}                                                        
\email{bharms@bama.ua.edu}
\affiliation{Department of Physics and Astronomy,                                   
The University of Alabama,
Box 870324, Tuscaloosa, AL 35487-0324}                                          
\begin{abstract}                                                                
The microcanonical ensemble is the proper ensemble to describe black holes
which are not in thermodynamic equilibrium, such as radiating black holes. 
This choice of ensemble eliminates the problems, e.g. negative specific heat
and loss of unitarity, encountered when the canonical ensemble is used.
In this review we present an overview of the weaknesses of the standard
thermodynamic description of black holes and show how the microcanonical
approach can provide a consistent description of black holes and their Hawking
radiation at all energy scales.
Our approach is based on viewing the horizon area as yielding the ensemble
density at fixed system energy.
We then compare the decay rates of  black holes in the two different pictures.
Our description is particularly relevant for the analysis of micro-black holes whose
existence is predicted in models with extra-spatial dimensions.                                                            
\end{abstract}                                                                  
\maketitle                                                                      
\section{Introduction}
Entropy as originally defined by Clausius is a statement about the
change of energy with respect to temperature for systems in
thermodynamic equilibrium. Later a correlation between entropy and
the statistical mechanical probability of finding a system in a
given state was obtained. The latter can be generalized to include
systems which are not in thermodynamic equilibrium and for which the
microcanonical ensemble is the proper ensemble to describe the
system. Since the microcanonical ensemble requires conservation of
energy for any system to which it is applied, it is the proper
ensemble to describe microscopic black holes. However, in spite of
its many mathematical and physical inconsistencies and drawbacks,
the treatment of black holes as thermodynamical systems has since
its inception been the description preferred by most physicists
investigating the nature of black holes. Not least among the
drawbacks is the fact that the laws of quantum mechanics are
violated, because the number density function of the emitted
radiation as calculated using a thermal vacuum is characteristic of
mixed states, while the incoming radiation may have been in pure
states. Since black holes can in principle radiate away completely,
the unitarity principle is violated.
\par
In a series of
papers~\cite{hl1_1,hl1_2,hl1_3,hl1_4,hl1_5,hl1_6,hl1_7,hl1_8,mfd} we
have investigated an alternative description of black holes which is
free of the encountered problems in the thermodynamical approach. In
this review, we shall first recall the main aspects of our approach
and then proceed to describe its application to microscopic black
holes~\cite{bhreview_1,bhreview_2,bwbh_1,bwbh_2,bwbh_3,bwbh_4,bwbh_5,
bwbh_6,bwbh_7,bwbh_8,bwbh_9,bwbh_10,bwbh_11,dadhich}. The latter are
predicted to exist in models with extra spatial
dimensions~\cite{add_1,add_2,add_3,RS}, and an extensive analysis of
their possible phenomenological appearance can be found in \cite{bhlhc_1,
bhlhc_2,bhlhc_3,bhlhc_4,bhlhc_5,bhlhc_6,
bhlhc_7,bhlhc_8,bhlhc_9,bhlhc_10,bhlhc_11,bhreview_1,bhreview_2,
chprd,chplb,string,CH,bhEarth1,bhEarth2,bulktidal,bwthick}.
\par
In Section~\ref{thermo} we present a brief summary of the
thermodynamical description of processes involving black holes and
discuss in detail the inconsistencies mentioned above; in
Section~\ref{Ddim} we briefly review Schwarzschild black holes in
more than four dimensions; In Section~\ref{mean} we discuss the
thermodynamical interpretation of black holes within the context of
mean field theory and show that the thermal vacuum is the false
vacuum for a black hole system. We also present an alternative
vacuum for such a system and the microcanonical number density which
corresponds to this vacuum. In Section~\ref{wave} we present the
microcanonical wave functions for the {\it in} and {\it out} states
and in Section~\ref{decay} we derive the black hole decay rates for
micro-black holes in models with extra dimensions.
\par
We shall either use natural units $c=\lp=\mpl=1$ or show explicitly
$\lp$ and $\mpl$ or the equivalent scales $\ld$ and $\md$ in models
with extra spatial dimensions. We shall also use the scale $M_{\rm
ew}\simeq 1\,$TeV.
\section{Thermodynamical Interpretation of Black Holes}
\label{thermo}
Bekenstein's original observation~\cite{bek_1,bek_2} that the area
of a black hole (in units of the Planck area $\lp^2$) is analogous
to the thermodynamical entropy, \be S_{\rm H} = \frac{A}{4} \
\label{S_H} \ee was enlarged upon in~\cite{bch} where the four laws
of black hole thermodynamics were hypothesized. The mass difference
of neighboring equilibrium states was shown to be related to the
change in the black hole area $A$ by the Smarr formula, \be \Delta M
= \kappa\,\Delta A +\varpi\,\Delta J+\Phi\,\Delta Q \ \ee where
$\kappa$ is the surface gravity related to the temperature by \be T
= \beta_{\rm H}^{-1}={\kappa\over{2\pi}} \ \label{betaH} \ee $J$ is
the angular momentum of the black hole, $Q$ its charge and $\varpi$,
$\Phi$ play the role of potentials. The partition function for the
black hole is assumed to be \be Z(\beta) = {\rm Tr}\, e^{-\beta\,
H}= e^{-S_{\rm H}} \ \label{part} \ee where $S_{\rm H}$ is the
Hawking entropy, \be S_{\rm H} =S_E- \beta_{\rm H}\,\varpi\,J \ \ee
and $S_E$ the Euclidean action. Finally, in thermodynamical
equilibrium the statistical mechanical density of states is given by
\be \Omega = Z^{-1}(\beta) = e^{S_{\rm H}} \; \ee and the specific
heat \be C_V = {\partial E\over{\partial T}} =
-{\beta^2\over{8\pi}}<0 \ \label{C<0} \ee which is a clear signal
that the thermodynamical analogy fails.
\par
A second problem can be best shown if we specialize the previous
expressions to the Schwarzschild black hole, for which $S_{\rm H} =
S_E= 4\,\pi\, M^2$ and \be \beta_{\rm H} = 8\,\pi\, M \ \ee It then
follows that the partition function as calculated from the
microcanonical density of states, \be Z(\beta) = \int_0^\infty dM\,
\Omega(M)\, e^{-\beta\, M} =\int_0^\infty dM\, e^{4\,\pi\, M^2}\,
e^{-\beta\, M} \to \infty \ \ee is infinite for all temperatures and
hence the canonical ensemble is not equivalent to the (more
fundamental) microcanonical ensemble, as is required for
thermodynamical equilibrium.
\par
The inequivalence of the two ensembles in systems with long-range
interactions, such as gravity, has been investigated extensively
(see, for example, the review~\cite{chavanis} for a presentation of
this issue in the astrophysical context). However, statistical
mechanical theorems show that the specific heat can be negative in
the microcanonical ensemble (where energy is held fixed) but must be
necessarily positive in the canonical ensemble (where temperature is
fixed). The above result~\eqref{C<0} therefore should be more
properly interpreted as implying that the microcanonical ensemble
can be used for a black hole, whereas the canonical ensemble is {\em
not well-defined\/} and should only be viewed as a useful
approximation as long as the temperature does not change
significantly on the time scale of interest.
\par
In fact, if quantum mechanical effects are taken into account, black
holes can be shown to evaporate and, according to the canonical
picture, the emitted radiation has a Planckian
distribution~\cite{hawk,gibb}, \be n_{\beta_{\rm H}}(\omega)={1\over
e^{\beta_{\rm H}\,\omega}-1} \ \label{n_b} \ee This implies that the
black hole mass $M$ will decrease in time, whereas the temperature
$\beta_{\rm H}^{-1}\sim M^{-1}$ will increase (possibly) without
bounds. Moreover, since in the standard Hawking's picture black
holes can in principle radiate away completely, this result implies
that information can be lost, because pure states can come into the
black hole but only mixed states come out. The breakdown of the
unitarity principle is one of the most serious drawbacks of the
thermodynamical interpretation, since it would require the
replacement of quantum mechanics with some new (unspecified)
physics. In order to avoid this scenario, some approaches predict
the evaporation leaves a ``black hole
remnant''~\cite{hoss_1,hoss_2}, as follows from assuming generalized
uncertainty principles~\cite{gup} and within models of
non-commutative geometry~\cite{piero}.
\section{Black holes in $D$ dimensions}
\label{Ddim}
The inconsistencies of the thermodynamical interpretation are an
indication that the interpretation of Equation~(\ref{part}) as the
canonical partition function is wrong. In analogy with the usual WKB
approximation, we instead hypothesize that \be P \simeq e^{-S_{\rm
H}} \ee is the probability for the transition from the metastable
vacuum, namely the black hole semiclassical vacuum, to the true
vacuum state with no black hole. This interpretation holds for any
kind of black hole. The picture of particles tunneling through the
horizon is technically not accurate, because the horizon is a causal
boundary, not a potential barrier. Such a potential might be
generated if the backreactions of the emitted particles are taken
into account~\cite{parikh_1,parikh_2}. However, the distribution of
emitted particles is decidedly non-thermal for such a potential.\
The quantum degeneracy of states for the system is proportional to
$P^{-1}$ and is then given by \be \sigma \simeq c\,e^{A/4} \
\label{sig} \ee where the constant $c$ is determined from quantum
field theoretic corrections and can contain non-local~effects.
\par
Explicit expressions can be obtained for the above quantities for
some geometries. We shall here just consider the $D$-dimensional
Schwarzschild black hole~\cite{mp}, \be ds^2 = -e^{2\,\lambda}\,
d\tau^2 + e^{-2\,\lambda}\, dr^2 + r^2\, d\Omega_{D-2}^2 \; \ee
where \be e^{2\,\lambda} = 1 - \left({r_{\rm
H}\over{r}}\right)^{D-3} \; \ee The horizon area in $D$ dimensions
is \be {A\over{4}} = {A_{D-2}\over{16\,\pi}}\, \beta_{\rm H}\,
r_{\rm H}^{D-2} \; \ee with \be M = {D-2\over{16\pi}}\, A_{D-2}\,
r_{\rm H}^{D-3} \; \ee where $A_{D-2}$ is the area of a unit
$(D-2)$-sphere. Eliminating the horizon radius $r_{\rm H}$ in favor
of the mass, the area becomes \be {A\over{4}} = C(D)\,
M^{D-2\over{D-3}} \; \ee where $C(D)$ is the mass-independent
function \be C(D) = {4^{D-1\over{D-3}}\, \pi^{D-2\over{D-3}}
\over{(D-3)(D-2)^{D-2\over{D-3}}\, A^{1\over{D-3}}_{D-2}}} \; \ee
From the above, we obtain the degeneracy of states \be \sigma(M)
\simeq \Omega^{-1}(M) \simeq c\,
\exp\left[C(D)\,M^{\frac{D-2}{D-3}}\right] \ \label{sigmaS} \ee
Comparing this expression to those known for non-local field
theories, we find that it corresponds to the degeneracy of states
for an extended quantum object ($p$-brane) of dimension $p =
{D-2\over{D-4}}$. As has been demonstrated by several authors
\cite{fub_1,fub_2,fub_3}, an exponentially rising density of states
is the clear signal of a non-local field theory. $P$-brane theories
are the only known non-local theories in theoretical physics which
can give rise to exponentially rising degeneracies.
\par
It is important to remark here that all dimensional quantities are
evaluated in units of the corresponding Planck scale. For example,
$M$ in Equation~(\ref{sigmaS}) is actually $M/M_{\rm P}$ in $D=4$
dimensions. In models with extra spatial dimensions, $M_{\rm P}\sim
10^{16}\,$TeV is replaced by a fundamental gravitational mass
$M_{(D)}$ which could be as low as $M_{\rm ew}\simeq 1\,$TeV. In the
following, we shall consider two such scenarios: the ADD
case~\cite{add_1,add_2,add_3} with compact extra dimensions, and the
RS case~\cite{RS} with one possibly very large extra~dimension.
\section{Quantum Field Theory on Black Hole Backgrounds}
\label{mean}
To study particle production and propagation in black hole
geometries we now turn to the mean field approximation in which
fields are quantized on a classical black hole background. Since
black holes have non-trivial topologies which causally separate two
regions of space, the number of degrees of freedom is doubled, and
two Fock spaces are required to describe quantum processes occurring
in the vicinity of a black hole. Calculations of quantities
associated with such processes can be carried out in ways analogous
to calculations in Thermofield Dynamics~\cite{umez}, but with an
overall fixed energy~\cite{mfd}.
\subsection{Canonical Formulation}
\label{TFD-f}
The thermal vacuum for quantum fields scattered off of black holes
can be written as \be |{\rm out};0\rangle =
{1\over{Z^{1/2}(\beta_{\rm H})}}\,\sum_{n=0}^{\infty}\,
e^{-\beta_{\rm H}\, n\,\omega/2}\,|n\rangle \otimes
|\tilde{n}\rangle \; \ee with the partition function \be Z(\beta) =
\sum_{n=0}^{\infty} e^{-\beta\, n\,\omega} \ee and the states
$|\tilde{n}\rangle $ are a complete orthonormal basis for the region
of space causally disconnected from an external observer. Operators
corresponding to physically measurable quantities are defined on the
basis set $|n\rangle $ for states outside the horizon. The ensemble
average (expectation value) of a physical observable $\hat O$ in the
$out$ region is \be \langle{\rm out};0|{\hat O}|{\rm out};0\rangle
={1\over{Z(\beta_{\rm H})}}\, \sum_{n} e^{-n\,\beta_{\rm
H}\,\omega}\langle n|{\hat O}|n\rangle \ \label{expec} \ee where the
temperature is given in Equation~(\ref{betaH}). For example, if
$\hat O$ is the number operator for particles of rest mass $m$, the
ensemble average given in Equation~(\ref{expec}) is the particle
number density~(\ref{n_b}).
\par
To describe particle interactions one then needs the (thermal)
particle propagator, \be \Delta_{\beta_{\rm H}} =
{1\over{k^2-m^2+i\epsilon}} - 2\,\pi\, i\, \delta(k^2-m^2)\,
n_{\beta_{\rm H}}(m,k) \; \label{delta} \ee with $n_{\beta_{\rm H}}$
given by Equation~(\ref{n_b}). These expressions are valid if black
holes are described by a local field theory. However, as discussed
in Section~\ref{thermo}, the particle number distribution given in
Equation~(\ref{n_b}) implies loss of coherence. The $in\/$ state is
a pure state \be |{\rm in};0\rangle  = |0\rangle  \otimes
|\tilde{0}\rangle \; \ee but the number density obtained from the
outgoing states is a thermal distribution.
\par
In the microcanonical approach non-local effects can be taken into
account by summing over all possible masses (angular momenta and
charges) \be n_{\beta_{\rm H}}(k) = \int_0^{\infty}dm\,
\sigma(m)\,n_{\beta_{\rm H}}(m,k) \; \ee Inclusion of non-local
effects changes the thermal vacuum to \be |{\rm out};0\rangle  =
{1\over{Z^{1/2}(\beta_{\rm H})}}
\left[\prod_{m,k}\,\sum_{n_{k,m}=0}^{\infty}\right]\,\prod_{m,k}
e^{-\beta_{\rm H}\, n_{k,m}\omega_{k,m}/2}\, |n_{k,m}\rangle \otimes
|\tilde{n}_{k,m}\rangle \ \ee where the quantity in square brackets
represents the product of the sums over the discrete values of the
momentum and mass. The canonical partition function extracted from
this expression is \be Z(\beta_{\rm H}) =
\exp\left(-{V\over{(2\pi)^{D-1}}}\int_{-\infty}^{+\infty}
d^{D-1}{\vec k}\,\int_0^{\infty} dm \, \sigma(m)\, \ln
[1-e^{-\beta_{\rm H}\,\omega_k(m)}]\right) \; \ee where the discrete
mass and momentum indices have been changed to continuous values. A
system in thermodynamical equilibrium must satisfy Hagedorn's
self-consistency condition \cite{hag_1,hag_2,hag_3}, \be
\int_0^{\infty}\Omega(E)\, e^{-\beta\,E_H}\, dE =
\exp\left(-{V\over{(2\,\pi)^{D-1}}}\,\int_{-\infty}^{+\infty}
d^{D-1}{\vec k}\, \int_0^{\infty}dm\, \sigma(m)\, \ln
[1-e^{-\beta_{\rm H}\,\omega_k(m)}] \right) \; \ee It is well known
that only strings ($p=1$) satisfy this condition \be \sigma(m) \sim
e^{b\; m} \; , \ \ \ \ (m \to \infty) \; \ee for $\beta_{\rm H} > b$
(Hagedorn's inverse temperature). But black holes are not strings,
as can be inferred from the quantum mechanical density of states for
Schwarzschild black holes~(\ref{sigmaS}). Therefore black holes do
not satisfy Hagedorn's condition (${D-2\over D-3}>1$ for $D>3$) and
are not in thermal equilibrium. We are thus led to conclude the
thermal vacuum is the false vacuum for a black hole and a better
description can be given by simply assuming energy conservation of
the entire black hole radiation system.
\subsection{Microcanonical Formulation}
The true vacuum for a black hole system of fixed total energy $E$
can be obtained by first writing the thermal vacuum in terms of the
density matrix $\hat\rho$ for a system in thermal equilibrium \be
|0(\beta)\rangle  = \hat{\rho}(\beta ; H) |\Im \rangle \; \ee where
\be \hat{\rho}(\beta,H) ={\rho(\beta,H)\over{\langle \Im|
\rho(\beta,H)|\Im\rangle }} \ \qquad {\rho} (\beta ; H) =
e^{-\beta\,H} \ee and the state \be |\Im\rangle  =
\left[\prod_{k,m}\sum_{n_{k,m}}\right]\,\prod_{k,m}|n_{k,m}\rangle
\otimes |\tilde{n}_{k,m}\rangle \; \ee The traces of observable
operators are given by \be {\rm Tr}\,{\hat O} = \langle \Im | {\hat
O}|\Im\rangle \; \ee For example the free field propagator can be
determined from \be \Delta_{\beta}^{ab} = -i\langle \Im| T
 \phi^a(x_1)\phi^b(x_2)\hat{\rho}|\Im\rangle
\; \ee The superscripts on $\phi$ refer to the member of the thermal
doublet~\cite{umez} \be \phi^a = \left(
\begin{matrix}
\phi \cr \tilde{\phi}^{\dagger}\cr
\end{matrix}
\right) \; \ee being considered. The Fourier transform of
$\Delta_{\beta}^{11}(x_1,x_2)$ (the physical component) is equal to
$\Delta_\beta$ given in Equation~(\ref{delta}).
\par
We can now formally define the microcanonical vacuum for an
evaporating black hole as \be |E\rangle  =
{1\over{\Omega(E)}}\int_0^E \Omega(E-E')\, L_{E-E'}^{-1}[|\beta_{\rm
H}\rangle ]\,dE' \; \ee where $L^{-1}$ is the inverse Laplace
transform. Using this basis, physical correlation functions are
expressed~as \be G^{a_1,...,a_N}_{E}(1,2...,N) = \langle
\Im|T\phi^{a_1}(1),...,\phi^{a_N}(N)|E\rangle \; \ee
\par
Interaction effects can be taken into account by means of the
microcanonical propagator \be \Delta_E^{11}(k) =
{1\over{k^2-m^2+i\,\epsilon}} - 2\,\pi\, i\,\delta\,(k^2-m^2)\,
n_E(m,k) \; \ee where $n_E(m,k)$ is the microcanonical number
density \be n_E(m,k) = \sum_{l=1}^\infty\,{\Omega(E-l\,\omega_k(m))
\over\Omega(E)}\,\theta(E-l\,\omega_k) \ \label{n_E} \ee which is
our candidate alternative to Equation~(\ref{n_b}) for the
distribution of particles emitted by a black hole. Strictly
speaking, this expression is valid only when the black hole system
is not too far from equilibrium. This expression is obtained by
taking the inverse Laplace transform of the canonical vacuum state
of thermofield dynamics~\cite{umez} and does not satisfy the
Principle of Equal Weights~\cite{kubo}. The general expression for
the microcanonical number density is given in \cite{leblanc}.
\section{Hawking Effect}
\label{wave}
The analysis carried on so far is global in nature. In fact,
although consistent equilibrium configurations for gases of black
holes and number densities for the emitted radiation in such
configurations can be found~\cite{mfd}, the geometry of spacetime
never appears explicitly in the final expressions. Of course, one is
also interested in the local properties of spacetime, and this is
most intriguing in the present case because the above results should
include implicitly any back-reactions of the radiation on the
metric. We then show that the wave functions in the microcanonical
vacuum can be obtained by making a formal replacement in the wave
functions obtained for the thermal vacuum.
\subsection{Thermal Vacuum}
In flat four-dimensional spacetime with spherical coordinates
$\{t,r,\theta,\phi\}$ incoming and outgoing spherical waves are
asymptotically given by \be &&\psi_{in} =
{Y_{lm}(\theta,\phi)\over{\sqrt{8\,\pi^2\,\omega}}}\,
{e^{-i\,\omega\, v}\over{r}} \ \qquad v = t+r_* \; \label{psiIn}
\\
&&\psi_{out} = {Y_{lm}(\theta,\phi)\over{\sqrt{8\,\pi^2\,\omega}}}\,
{e^{-i\,\omega\, u}\over{r}} \ \qquad u = t-r_* \ \label{psiOut} \ee
If we now consider waves propagating on a Schwarzschild black hole,
and do not take into account back-reactions, the incoming wave
becomes~\cite{birr} \be \psi_{in} = \left\{
\begin{matrix}
\strut\displaystyle{ {Y_{lm}(\theta,\phi)
\over{\sqrt{8\,\pi^2\,\omega}}}\,
{e^{i\,(\omega/\kappa)\,\ln(v_0-v)}\over{r}}} & \quad v < v_0\cr \cr
0 & \quad v > v_0 \ \cr
\end{matrix}
\right. \label{psi_T} \ee which obeys the wave equation in a
background with surface gravity $\kappa$. The $in\/$ states for the
two vacua are related by the Bogolubov transformation \be
\left.\begin{matrix} \alpha_{\omega\omega'} \cr \cr
\beta_{\omega\omega'}\cr
\end{matrix}
\right\} = {1\over{2\,\pi}} \int_{-\infty}^{v_0} dv
\,\left({\omega'\over{\omega}}\right)^{1/2} e^{\pm
i\,\omega'\,v}\,e^{i\,(\omega/\kappa)\,\ln[(v_0-v)/c]} \
\label{bogo} \ee where $c$ is a constant. The two coefficients
$\alpha$ and $\beta$ are related by the Wronskian condition \be
\sum_{\omega'}\left(|\alpha_{\omega\omega'}|^2 -
|\beta_{\omega\omega'}|^2\right) = 1 \ \label{omeg} \ee The
integrals in Equation~(\ref{bogo}) can be evaluated explicitly, \be
|\alpha_{\omega\omega'}|^2 = e^{2\,\pi\,\omega/\kappa}\,
|\beta_{\omega\omega'}|^2 \ \label{ab} \ee which substituted into
Equation~(\ref{omeg}) yields the Planckian distribution~(\ref{n_b}).
\subsection{Microcanonical Vacuum}
The relationship between $\alpha$ and $\beta$ in Equation~(\ref{ab})
arises because the logarithmic term in Equation~(\ref{bogo})
introduces a branch cut, and the integration around this branch cut
causes the factor multiplying this term (times $2\pi$) to appear in
the exponential multiplying $\beta$. Thus if we simply make the
formal replacement \be {2\,\pi\,\omega\over{\kappa}} \to
\ln(1+n_E^{-1}(\omega)) \; \label{replace} \ee where $n_E(\omega)$
is the microcanonical number density~(\ref{n_E}), the {\em out\/}
waves are of the form~(\ref{psiOut}) and \be \psi_{in} = \left\{
\begin{matrix}
\strut\displaystyle{ {Y_{lm}(\theta,\phi)
\over{\sqrt{8\,\pi^2\,\omega}}}\,
{e^{(i/2\,\pi)\,\ln[1+n_E^{-1}(\omega)]\,\ln(v_0-v)}\over{r}}} &\ \
\ v < v_0\cr \cr 0 &\ \ \ v > v_0 \
\end{matrix}
\right. \label{psi} \ee The relation between $\alpha$ and $\beta$
now becomes \be |\alpha_{\omega\omega'}|^2 = e
^{\ln(1+n_E^{-1}(\omega))}\, |\beta_{\omega\omega'}|^2 \; \ee which
gives for the sum over $\omega'$ \be
\sum_{\omega'}|\beta_{\omega\omega'}|^2 = n_E(\omega) \ \ee The wave
in Equation~(\ref{psi}) does not satisfy the same equation as the
wave in Equation~(\ref{psi_T}), but it should satisfy a wave
equation in a background whose metric includes back-reaction and
non-local effects.
\section{Micro-Black Hole Decay Rates}
\label{decay}
As an example of the differences between the predictions of the two
approaches (thermal vs.~microcanonical) suppose we consider
$D$-dimensional Schwarzschild black holes. In order to allow for the
cases with extra spatial dimensions, it will be convenient to show
all fundamental constants explicitly. For example, the
microcanonical number density~(\ref{n_E}) is given by \be
n_{(D)}=B\sum_{n=1}^{[[M/\omega]]}\, \exp \left\{\frac{S_{(D)}^{\rm
E}(M-n\,\omega)}{\lp\,\mpl} -\frac{S_{(D)}^{\rm
E}(M)}{\lp\,\mpl}\right\} \ \label{n_M} \ee where $S_{(D)}^{\rm E}$
is the Euclidean action, $[[X]]$ denotes the integer part of $X$ and
$B=B(\omega)$ encodes deviations from the area law (in the following
we shall assume $B$ is constant in the range of interesting values
of $M$). From Equation~(\ref{sigmaS}), we immediately obtain \be
\frac{S_{(D)}^{\rm E}(M)}{\lp\,\mpl} =
\left(\frac{M}{\meff}\right)^\delta \equiv \tilde M^\delta \
\label{areaL} \ee This means the black hole degeneracy is counted in
units of $\meff=\md$ and the luminosity \be {\cal L}_{(D)}(M) =
\int_0^\infty \sum_{s=1}^S n_{(D)}(\omega)\,
\Gamma_{(D)}^{(s)}(\omega)\,\omega^{D-1}\,\ud\omega \ \ee becomes
\be {\cal L}(M) = B\,e^{-\tilde M^\beta}\,
\int_0^\infty\sum_{n=1}^{[[\tilde M/\tilde\omega]]} e^{\left({\tilde
M-n\,\tilde\omega}\right)^\delta}
\tilde\omega^{D-1}\,\ud\tilde\omega \ \label{LH} \ee For the
four-dimensional Schwarzschild black hole, $\delta=2$, $n(\omega)$
mimics the canonical ensemble (Planckian) number density in the
limit $M \to \infty$, and the luminosity becomes \be {\cal L}_{\rm
H} \sim \int_0^\infty\frac{\omega^{3}\,\ud\omega}{e^{\beta_{\rm
H}\,\omega}\mp 1} \sim T_{\rm H}^{4} \ \ee where $T_{\rm
H}=\beta_{\rm H}^{-1}=1/(8\,\pi\,M)$ is the Hawking temperature.
Upon multiplying by the horizon area, one then obtains the Hawking
evaporation rate~\cite{hawk} \be \frac{\ud M}{\ud\tau} \simeq
\frac{\geff\,\mpl^3}{960\,\pi\,\lp\,M^2} \ \label{eva4} \ee where
$\geff\simeq 10-100$ is the number of effective degrees of freedom
into which a four-dimensional black hole can evaporate, usually
assumed equal to the number of Standard Model particles plus
gravitational modes with energy smaller than the instantaneous black
hole temperature $\beta_{\rm H}^{-1}$~\cite{mac}.
\subsection{ADD Scenario}
If the space-time is higher dimensional and the $d=D-4$ extra
dimensions are compact and of size $L$, the relation between the
mass of a spherically symmetric black hole and its horizon radius is
changed to~\cite{mp,bhlhc_1,bhlhc_2,bhlhc_3,bhlhc_4,bhlhc_5,bhlhc_6,
bhlhc_7,bhlhc_8,bhlhc_9,bhlhc_10,bhlhc_11} (For the microcanonical
description of micro-black holes in the ADD scenario, see
also~\cite{hoss_1,hoss_2,kotwai}.) \be R_{\rm H}\simeq
\ell_{(4+d)}\, \left({2\,M\over M_{(4+d)}}\right)^{1\over 1+d} \
\label{R_H<} \ee where $G_{(4+d)}\simeq L^d\,G_{\rm N}$ is the
fundamental gravitational constant in $4+d$ dimensions.

\begin{figure}[t]
\centering \raisebox{4cm}{${\mathcal L}_{(10)}$} \epsfxsize=3.5in
\epsfbox{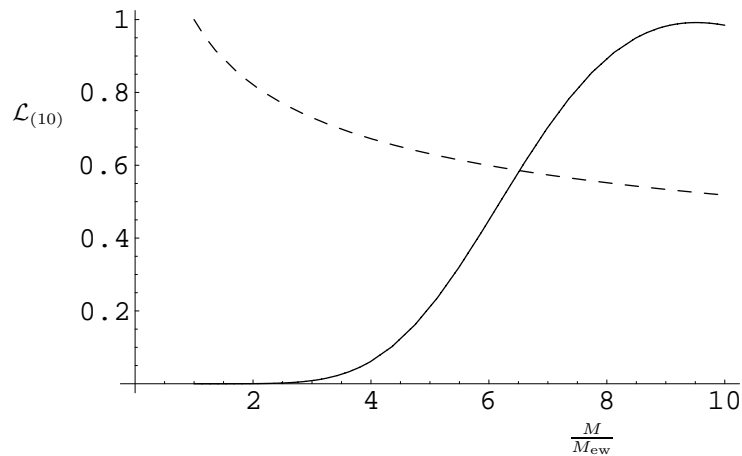}
\\
\hspace{-0.2in} \raisebox{0.5cm} {\hspace{6cm} ${M\over M_{\rm
ew}}$}
\caption{Microcanonical luminosity (solid line) for a small black
hole with $d=6$ extra dimensions compared to the corresponding
canonical luminosity (dashed line). Vertical units are chosen such
that ${\mathcal L}^H_{(10)}(M_{\rm ew})=1$.} \label{L_add}
\end{figure}

The Euclidean action is of the form in Equation~(\ref{areaL}) with
$\meff=M_{(4+d)}\sim M_{\rm ew}$ and \linebreak
$\delta=(d+2)/(d+1)$. In four dimensions one knows that
microcanonical corrections to the luminosity become effective only
for $M\sim \mpl$, therefore, for black holes with $M\gg M_{\rm ew}$
the luminosity~(\ref{LH}) should reduce to the canonical result. In
order to eliminate the factor $B$ from Equation~(\ref{LH}), one can
therefore equate the microcanonical luminosity to the canonical
expression at a given reference mass $M_0\gg M_{\rm ew}$ and then
normalize the microcanonical luminosity according to \be {\mathcal
L}_{(4+d)}(M)&\simeq& {{\mathcal L}_{(4+d)}^H(M_0) \over {\mathcal
L}_{(4+d)}(M_0)}\, {\mathcal L}_{(4+d)}(M) \ \label{L_norm} \ee The
black hole luminosity thus obtained differs significantly from the
canonical one for $M\sim M_{\rm ew}$, as can be clearly seen from
the plot for $d=6$ in Figure~\ref{L_add}. For smaller values of $d$
the picture remains qualitatively the same, except that the peak in
the microcanonical luminosity shifts to lower values of $M$.
Although the integral in Equation~(\ref{LH}) can now be performed
exactly, its expression is very complicated and we omit it. In all
cases, the microcanonical luminosity becomes smaller for $M\sim
M_{\rm ew}$ than it would be according to the canonical luminosity,
which makes the life time of the black hole longer than in the
canonical picture~\cite{CH}.
\vspace{-24pt}
\subsection{RS Scenario}
In order to study this case, we shall make use of the solution given
in ~\cite{dadhich}, \be \ud s^2 = - A\,\ud t^2 + A^{-1}\,\ud r^2 +
r^2\,\ud\Omega^2 \ \label{tidal} \ee with \be
A=1-\frac{2\,\lp\,M}{\mpl\,r}-q\,\frac{\mpl^2\,\lp^2}{\mdd^2\,r^2} \
\ee where $q$ is the so called tidal charge. For $q>0$, this metric
has one horizon at \be
\rh=\lp\left(\frac{M}{\mpl}+\sqrt{\frac{M^2}{\mpl^2}+q\,\frac{\mpl^2}{\mdd^2}}\right)
\ \label{rhq} \ee It is then plausible that both the mass $M$ and
the (dimensionless) tidal charge $q$ depend upon the black hole
proper mass $M_0$ in such a way that when $M_0$ vanishes, so do $M$
and $q$. The functions $M=M(M_0)$ and $q=q(M_0)$ could only be
determined precisely by solving the full bulk equations, for example
using the four-dimensional metric~\eqref{tidal} as a boundary
condition. Unfortunately, this task cannot be performed exactly, but
only numerically or
perturbatively~\cite{bwbh_1,bwbh_2,bwbh_3,bwbh_4,bwbh_5,
bwbh_6,bwbh_7,bwbh_8,bwbh_9,bwbh_10,bwbh_11}.
\par
In order to simplify the analysis, we shall first assume that
$M=M_0$ and, at least for $M\sim \mdd$, that the functional form of
$q$ is given by \be q \simeq
\left(\frac{\mpl}{\mdd}\right)^\alpha\left(\frac{M}{\mdd}\right)^\beta
\ \ee where $\alpha$ and $\beta> 0$ are real parameters. The
luminosity~(\ref{LH}) can then be computed exactly and a complete
survey is given in ~\cite{bhEarth1,bhEarth2,bwthick}. In general,
the decay rate is well-approximated by a power law,~namely \be
\left.\frac{\ud M}{\ud\tau}\right|_{\rm evap} \simeq C\,M^s \
\label{evax} \ee where $s$ can be determined analytically for
special cases and numerically in general~\cite{bhEarth2}.
\par
For instance, let us work out the case with $\beta=1$ and
$\delta=1$. The effective four-dimensional Euclidean action is given
by \be \frac{S_{(4)}^{\rm E}}{\lp\,\mpl} \simeq \frac{M}{\meff} \
\label{SE} \ee with \be \meff =4\,\mpl\,\frac{\lp}{L} \ \ee and the
luminosity in this case is simple enough, that is \be {\cal L}
\simeq B\,e^{-\tilde M}\, \sum_{n=1}^{\infty}\frac{1}{n^4}
\int_0^{\tilde M} e^{x}\left(\tilde M-x\right)^3 \ud x \simeq \tilde
B \ \label{Lq} \ee where we used $\meff\ll\mew\sim M$ and $\tilde B$
is a new constant. Upon multiplying by the horizon area, we then get
the microcanonical evaporation rate per unit proper time \be
\left.\frac{\ud M}{\ud\tau}\right|_{\rm evap} \simeq C\,M \
\label{eva5} \ee where $C$ is again a constant we can determine by
equating the rate~\eqref{eva5} with the Hawking
expression~\eqref{eva4} for $M=\mc$ defined by $\rh(\mc)\simeq L$.
\section{Summary}
We have reviewed the main arguments in favor of the microcanonical
ensemble for describing evaporating black holes. A shortcoming of
the foregoing analysis is the use of the mean field approximation.
However, all calculations of particle emission utilize this
approximation, and the microcanonical approach is clearly preferable
to the thermodynamical approach in the semiclassical quantization
processes described above. It is free of the inconsistencies present
in the thermodynamical approach, and its predictions seem to be more
physically reasonable, e.g.,~a finite black hole decay rate
throughout the life of the black hole. The use of a fixed energy
basis for the Hilbert space of the theory instead of the usual
thermal state implies that black holes are particle states. In our
interpretation of black holes as quantum objects the associated
quantum degeneracy of states obtained from the inverse of the
tunneling probability points to the identification of black holes
with the excitation modes of $p$-branes.
\par
For a four-dimensional black hole the above picture leads to very
small, undetectable, departures from the usual Hawking picture.
However, if extra dimensions exist, and the fundamental scale of
quantum gravity is as low as $1\,$TeV, microscopic black holes with
a mass of a few TeV's might be produced in modern accelerators. In
this case the microcanonical description then becomes a necessary
tool to describe their evaporation, and there is no need for the
thermodynamical concept of entropy for microscopic black holes. In
general, one then expects an increased life time with respect to
what would be predicted by the canonical ensemble, with the ending
stage of the evaporation resembling a more conventional,
quasi-exponential decay. For more details on the phenomenological
signatures of microscopic black holes at the LHC in the
microcanonical treatment, we just refer the reader to
\cite{hoss_1,hoss_2,ging}.
%
%




\bibliographystyle{mdpi}

\makeatletter

\renewcommand\@biblabel[1]{#1. }

\makeatother

%
%
%

\end{document}